\documentclass[]{article}

\begin{document}
\title{On the relation between quantum mechanical probabilities and event frequencies}
\author{Charis Anastopoulos \thanks{anastop@physics.upatras.gr}, \\
Department of Physics, University of Patras, \\
26194 Patras, Greece } \maketitle
\begin{abstract}
The probability `measure' for  measurements at two consecutive
moments of time is non-additive. These probabilities, on the other
hand,  may be determined by the limit of relative frequency of
measured events, which are by nature additive.
 We demonstrate that there are only two ways to resolve
this problem.
 The first solution places emphasis on the precise use of the concept of conditional probability
 for successive measurements. The physically correct
conditional probabilities define additive probabilities for
two-time measurements. These probabilities depend explicitly on
the resolution of the physical device and do not, therefore,
correspond to a function of the associated projection operators.
It follows that quantum theory distinguishes between physical
events and propositions about events, the latter are not
represented by projection operators and that the outcomes of
two-time experiments cannot be described by quantum logic.

The alternative explanation is rather radical: it is conceivable
that the relative frequencies for two-time measurements do not
converge, unless a particular consistency condition is satisfied.
If this is true, a strong revision of the quantum mechanical
formalism may prove necessary. We stress that it is possible to
perform experiments that will distinguish the two alternatives.

\end{abstract}

\renewcommand {\thesection}{\arabic{section}}
 \renewcommand {\theequation}{\thesection. \arabic{equation}}
\let \ssection = \section
\renewcommand{\section}{\setcounter{equation}{0} \ssection}

\section{Introduction}

Quantum mechanics is a probabilistic theory. It provides a set of
rules that allows us to  associate probabilities to specific
physical events. There is little doubt that these rules have been
proved remarkably successful in the description of any physical
phenomenon that we have been able to study experimentally, but
gravitational ones.

The reasoning in terms of probability in physical theories,
however, is not entirely unproblematic. Most  discussions about
the interpretation of quantum theory -- and the associated
"paradoxes"-- are related to the appropriate use of quantum
probabilities. Some of the issues raised are not specifically
quantum -- they refer to the physical applicability of the general
concepts of probability theory and date at least back to
Boltzmann.

One may ask, for instance, whether the probabilities are
subjective or objective-- namely whether they refer to our
knowledge about a physical system or to the physical system
itself. In the latter case, one may further ask whether
probabilities refer to an individual system --denoting perhaps its
propensity to manifest one behavior or another- or to statistical
ensembles. One may also question whether there exists a sample
space for quantum phenomena, or all predictions have to make
reference to a concrete measurement set-up.

There exists a common denominator in all interpretations of
probability, either classical or quantum. We may not agree whether
probabilities refer to the properties of the things themselves or
not , but we do accept that probabilities refer to the statistics
of measurement outcomes. Probabilities may or may not be
physically meaningful {\em a priori}  (before an experiment), but
they can definitely be determined {\em a posteriori}, namely after
a large number of experimental runs.

The probability of an event is defined as the limit of the
relative frequency of this event as the number of trials goes to
infinity. It may be argued that this is not the only way we employ
probability in physics -- after all statistical arguments enter
into the design and preparation of any experiment. Still, whenever
we want to compare the theoretical probabilities with concrete
empirical data, we invariably employ the relation of probability
to event frequencies.

In this paper we analyse the basic properties of quantum
mechanical probability for two-time measurements in two
consecutive moments of time (two-time measurements). The key point
of our argumentation is the empirical determination of
probabilities as limits of relative frequencies. We employ this
relation without committing to a frequency interpretation of
probability \cite{vMis}-- we need not assume that probability, as
a concept,  is {\em defined} as a limit of relative frequencies.
Neither do we commit to a specific interpretation of quantum
theory. We only assume that the outcomes of measurements (that
have actually been performed) are described by the probabilities
obtained from the rules of quantum theory. Quantum mechanical
probabilities may refer to other aspects of physical reality, but
we need not make such an assumption. This thesis can hardly be
rejected by any interpretation of quantum theory.

In a two-time measurement one determines specific properties of a
physical system at two successive moments of time. The measurement
outcomes may be sampled in the same manner they are sampled in the
single-time measurements. Probabilities are then still determined
by the limits of relative frequencies. We may still employ the
rules of quantum theory to associate a probability to each
possible measurement outcome. The problem is that the quantum
mechanical probability `measure' for two-time histories does not
satisfy the additivity property of probabilities. On the other
hand, relative frequencies are always additive, since they are
constructed by counting specific and indivisible physical events.

We next proceed to resolve this conflict. A physical theory must
explain the observed phenomena -- namely the frequencies of
measured events. If these frequencies define probabilities, we
have to accept that the conventional rule for probabilities of
two-time measurements fails. We show that the derivation of this
rule employs the concept of conditional probability in a rather
ambiguous way. We address this problem and define thereby additive
probabilities for two-time measurements. But the new probability
assignment depends explicitly on the resolution of the physical
device. The probabilities assigned to a specific sample set of
measurements depend therefore on the physical characteristics of
the apparatus and {\em they are not a function of the associated
projection operators}. Projection operators cannot represent
events universally. It follows that the YES-NO experiments
\cite{Jauch} cannot reconstruct all probabilistic aspects of a
physical system, and for this reason the outcomes of two-time
experiments cannot be represented by any form of quantum logic.

The alternative is rather radical, but cannot be {\em a priori}
rejected. It is conceivable that the relative frequencies for the
two-time measurements do not converge (see \cite{Khr00} for a
relevant interpretation of quantum probability ). In that case,
probabilities can only be defined for two-time events that satisfy
a consistency condition --the same condition that appears in the
consistent histories interpretation of quantum theory
\cite{Gri84,Omn8894, GeHa9093, Har93a, I94}. The failure of the
frequencies to converge is eventually due to the interference
between the two alternatives. This alternative explanation
implies, of course, that one would need an all-new reformulation
of quantum theory.

It is important to emphasise that the two possible resolutions of
our problem may be empirically distinguished. It is possible -- in
theory and we believe in practice too--  to design experiments
that will determine whether the relative frequencies of two-time
events converge or not. Either way, such experiments would shed
much light in many counter-intuitive aspects of quantum
probability.

\section{Probabilities for two-time measurements}
First we describe the relation of probabilities to event
frequencies. We assume an ensemble of a large number of
identically prepared systems. In each system we measure some
physical properties say $A$, which take value in a set $\Omega$.
We then perform the measurements one by one -- thus constructing a
sequence $A_N$ of points of $\Omega$, where $N$ is an integer that
labels the experiments. We next sample the measurement outcomes
into subsets $U$ of $\Omega$\footnote{ In standard probability
theory such sets are denoted as {\em events}, because they
represent possible outcomes of the measurement process. Not any
subset of $\Omega$ can play that role -- unless $\Omega$ is a
denumerable set. In the standard theory $U$ are usually taken to
be Borel sets, namely sets that can be obtained from denumerable
unions and intersections of the open subsets of $\Omega$. The
description of probabilities in terms of frequencies, however, is
related directly to sampling of measurement outcomes. If $\Omega$
is a continuous set, Borel subsets consisting of discrete points
are irrelevant to what we may actually measure and should be,
therefore, excluded.}.

We next define $n(U,N)$ as the number of times  that the result of
the measurement is found in $U$ in the first $N$ experiments. It
is evident that $n(U,N)$ satisfies the following properties
\begin{eqnarray}
n(U,N) &\geq& n(U,M), \, \mbox{if} \, N > M \\
n(U \cup V,N) &=& n(U,N) + n(V,N), \, \mbox{if}\,  U \cap V = \emptyset\\
n(\Omega,N) &=& N \\
n(\emptyset,N) &=& 0
\end{eqnarray}

One then may define the {\em a-posteriori} probability that an
event in $U$ has been realised as
\begin{eqnarray}
p(U) = \lim_{N \rightarrow \infty} \frac{n(U,N)}{N},
\end{eqnarray}
provided the limit exists. These probabilities satisfy all axioms
of ordinary probability.

We then consider a measurement at two successive moment of time.
At time $t=0$ a particle described the density matrix
$\hat{\rho}_0$ is emitted from a source. At time $t_1$ it passes
through a  device, which measures its position. The device may
simply consist of a thin strip of a medium, which registers a
track of the particles that penetrate it. The particle then passes
at time $t_2
> t_1$ from an identical measurement device, which we placed
immediately behind the first. It leaves a second track there.

The experiment described above may be repeated $N$ times, each
time recording the mark left by the particle on the measurement
devices. To study the statistics of the measurements, we split the
possible values of particle position at each moment of time into a
set of $n$ exclusive alternatives. Each alternative labelled by
the index $i$ corresponds to a subset $U_i$ of the real line, such
that $\cup_i U_i = {\bf R}$ and $U_i \cap U_j = \emptyset$, for $i
\neq j$.

The sample space $\Omega = {\bf R}^2$ for the two-times
measurements is partitioned into the $n^2$  sets $U_{ij} = U_i
\times U_j$, which are labelled by the ordered pair of integers
$(i,j)$. From the results of the measurements we may immediately
read the numbers $n(U_{ij},N)$. According to the relation between
relative frequencies and probabilities these numbers should
satisfy
\begin{eqnarray}
\frac{n(U_{ij},N)}{N} \rightarrow p(U_i,t_1;U_j,t_2),
\end{eqnarray}
as $N \rightarrow \infty$. Here $p(U_i,t_1;U_j,t_2)$ refers to the
probability first $i$ and then $j$ are realised.

The rules of quantum theory allow us to express this probability
in terms of the projection operators $\hat{P}_i$ that correspond
to the interval $U_i$ of the particle's position.
\begin{eqnarray}
p(U_i,t_1; U_j, t_2) = Tr(\hat{Q}_j\hat{P}_i\rho(t_1)\hat{P_i}),
\end{eqnarray}
where we denoted for simplicity $Q_j = e^{i\hat{H}(t_2-t_1)} P_j
e^{-i\hat{H}(t_2-t_1)}$.  $\hat{H}$ the Hamiltonian of the
particle and $\rho(t_1)$ the initial density matrix evolved until
time $t_1$.

Suppose, however, that we want to consider the probability that
the particle first crossed through either $U_1$ or $U_2$ and then
through $U_j$. The projection operator corresponding to $U_1 \cup
U_2 $ is $\hat{P_1} + \hat{P}_2$, hence the corresponding
probability is
\begin{eqnarray}
p(U_1 \cup U_2, t_1; U_j, t_2) = Tr(\hat{Q}_j(\hat{P}_1 +
\hat{P}_2) \hat{\rho}(t_1)(\hat{P}_1 + \hat{P}_2))\nonumber \\ =
p(U_1, t_1; U_j, t_2) + p(U_2, t_1; U_j, t_2) + 2 Re \, d(U_1,
U_2, t_1; U_j, t_2),
\end{eqnarray}
where
\begin{eqnarray}
d(U_1, U_2, t_1; U_j, t_2) = Tr ( \hat{Q}_j \hat{P}_1
\hat{\rho}(t_1) \hat{P}_2)
\end{eqnarray}
is known as the {\em decoherence functional} in the consistent
histories approach. It provides a measure of the interference
between the events 1 and 2.

On the other hand  the elementary properties of the frequencies
$n(U_{ij}, N)$ state that $ n([U_1 \cup U_2]\times U_j,N) = n(U_1
\times U_j,N) + n(U_2 \times U_j,N)$, so that in the limit $N
\rightarrow \infty$
\begin{eqnarray}
p(U_1 \cup U_2, t_1;  U_j, t_2) = p(U_1, t_1;, U_j, t_2) + p(U_2,
t_1; U_j, t_2)
\end{eqnarray}
In other words, the quantum mechanical probabilities are not
additive (unless the consistency condition $Re d(U_1, U_2, t_1;
U_j, t_2) = 0$ is satisfied), while the measured frequencies of
events are additive. We shall see that there exist only two
possible resolutions to the problem. The first one is close to
conventional wisdom about quantum theory, but has, nonetheless,
disturbing implications. The other is more radical, but cannot be
discounted {\em a priori}.

\section{Probabilities are contextual}

\subsection{The correct use of conditional probability}
Our first alternative involves the assumption that the sequences
(2.6) converge, while the second that they do not converge. In the
former case the physically relevant probabilities are defined by
the limit of the relative frequencies. These probabilities are a
datum of experiment, and as such they should be explained by the
physical theory. If the theory fails in that regard, then there
must be a mistake somewhere in the analysis. It follows that if
the probabilities can be  defined, the derivation of equation
(2.7) should be reexamined.

We start from a density matrix $\hat{\rho}$ at $t=0$, which is
evolved unitarily until time $t_1$, when the particle enters the
measuring device. If we register the particle in the interval
labelled by $i$, the outcoming density matrix will equal
\begin{eqnarray}
\frac{\hat{P}_i \hat{\rho}(t_1) \hat{P}_i}{Tr (\hat{\rho}(t_1)
\hat{P}_i)}.
\end{eqnarray}
We need make no commitments about the interpretation of the
measurement process. It is irrelevant whether the measuring device
is classical like in Copenhagen quantum theory, or quantum
mechanical and a physical process of wave packet reduction has
taken place. And it makes little difference whether the density
matrix refers to an individual system, or a statistical ensemble,
because at the end of the day our results will be interpreted by
statistical processing of the measurement outcomes. What is
important is that the density matrix (3.1) allows us to compute
the conditional probabilities that the event $j$ takes place at
$t_2$ provided the event $i$ took place at $t_1$
\begin{eqnarray}
\frac{Tr \left(\hat{P}_i \hat{\rho}(t_1) \hat{P}_i \hat{Q}_j
\right)}{Tr (\hat{\rho}(t_1) \hat{P}_i)},
\end{eqnarray}
from which the classical definition of conditional probability
leads us to expression (2.7) for the probability that first the
event $i$ takes place at $t_1$ {\em and then} the event $j$ takes
place at time $t_2$.

The problem lies in equation (3.1). If, instead of sampling the
measurement outcomes in the set, say $U_1$, we sampled it into
$U_1 \cup U_2$, we would have employed the projector $\hat{P}_1 +
\hat{P}_2$ and the out-coming density matrix would read
\begin{eqnarray}
\frac{(\hat{P}_1 + \hat{P}_2)\hat{\rho}(t_1) (\hat{P}_1 +
\hat{P}_2)}{Tr (\hat{\rho}(t_1) (\hat{P}_1 + \hat{P}_2)))}.
\end{eqnarray}
We would then obtain the result (2.8), which is inconsistent with
the probabilities defined through relative frequency.

However, there is no {\em a priori} reason to use equation (3.1)
for the out-coming density matrix. The action of the projection
$\hat{P}_i$ depends on our choice of sampling of measurement
outcomes and not on the measurement outcome itself. What has
actually taken place is that the particle left a mark on a
specific point, and we then {\em choose} to place that point into
one or the other set. If we had a measurement at a single moment
of time, this would not have been a problem, because the density
matrix (3.1) does not appear in any physical predictions for
single-time measurements. In a single-time measurement the only
physically relevant quantities are the probabilities $Tr(\hat{P}_i
\hat{\rho}(t_1))$, which are additive and for this reason they do
not depend on our choice of sampling.

In the two-time measurement, however, the  probabilities turn out
to be non-additive, and this should urge some caution on the use
of conditional probability. If we sample events into larger sets
than the ones being manifested in the experiments, then we employ
less information than what we have actually obtained.  Our use of
conditional probabilities will be, therefore, improper. (See the
discussion in \cite{Jaynes} about the way conditional
probabilities may lead to erroneous predictions, if we fail to
make use of all available information.)

 The physically correct
procedure would be to incorporate in our probabilities {\em all}
information that has been obtained from the measurements. In other
words we must construct the out-coming density matrix not on the
basis of our arbitrary choice of sampling events, but on what we
have actually observed. We should not use an arbitrary set of
projectors, but only the finest possible projectors compatible
with the resolution of the apparatus. If $\delta$ is the sharpest
resolution of the measuring device (say the width of the dots
indicating the particle's position) the relevant projectors are
$\hat{P}^{\delta}_x$, which project onto the interval $[x-
\frac{\delta}{2}, x+ \frac{\delta}{2}]$. Using these projectors we
construct the probabilities
\begin{eqnarray}
p_{\delta}(x_1,t_1;x_2, t_2) = Tr \left( e^{i\hat{H}(t_2-t_1)}
\hat{P}^{\delta}_{x_2} e^{-i\hat{H}(t_2-t_1)}
\hat{P}^{\delta}_{x_1} \hat{\rho}(t_1) \hat{P}^{\delta}_{x_1}
\right),
\end{eqnarray}
that a dot will be found centered at the point $x_1$ in the first
measurement and then a dot centered at the point $x_2$ in the
second measurement.

We may then construct the probabilities for a particle to be found
within a subset $U_i$ of ${\bf R}$ at time $t_1$ and then within a
subset $U_j$ at time $t_2$. For this purpose, we split each set
$U_i$ into mutually exclusive cells $u_{\alpha i}$ of size
$\delta$, such that
\begin{eqnarray}
\cup_{\alpha} u_{\alpha i} = U_i \\
u_{\alpha i} \cap u_{\beta i} = \emptyset, \alpha \neq \beta.
\end{eqnarray}
If we denote  select points $x_{\alpha i} \in u_{\alpha i}$, for
all $i$ ($x_{\alpha i}$ may be the midpoint of $u_{\alpha i}$), we
may construct
 the probability $p_{\delta}(U_i,t_1;U_j,t_2)$
 \begin{eqnarray}
p_{\delta}(U_i,t_1;U_j,t_2) = \sum_{\alpha} \sum_{\beta}
p_{\delta}(x_{\alpha i},t_1 ; x_{\beta j}, t_2)
 \end{eqnarray}

In the limit that the typical size of the sets $U_j$ is much
larger than $\delta$, we may approximate the summation by an
integral,

\begin{eqnarray}
p_{\delta}(U_i,t_1|U_j, t_2) = \frac{1}{\delta^2} \int_{U_i} dx_1
\int_{U_j} dx_2 p_{\delta}(x_1,t_1;x_2, t_2),
\end{eqnarray}
In other words, the objects
$\frac{1}{\delta^2}p_{\delta}(U_i,t_1|U_j, t_2)$ play the role of
probability densities. The probabilities (3.8) are compatible with
the relative frequencies, because they do satisfy the additivity
criterion. Note, however, that they depend strongly on the
resolution $\delta$ of the measuring device\footnote{If the two
-time probabilities did not depend on $\delta$, it would have been
possible to describe the quantum mechanical system in terms of a
stochastic process that reproduces all $n$-point functions of the
quantum mechanical description without making any reference to the
measurement apparatus. The generic dependence of $n$-time
probabilities on the measuring device renders this impossible, in
agreement with many constraints placed by Bell's theorem}.

There exists a systematic error in the definition of the
probabilities (3.8), which is due to the approximation of the sums
(3.7) by integrals. This is  related to the fact that the dots
have a finite size and hence cannot be definitely ascertained
whether they lie in a sample set $U_i$ or its neighboring one. For
sufficiently large sets $U_j$, characterised by a typical size
$L$, the ambiguity may be approximated by a Gaussian distribution
and is of the order $e^{-L^2/\delta^2}$.

The important feature of the probabilities (3.8) is that {\em they
are not functions of the projection operators $P_{U_i}$}. They
depend on the resolution $\delta$ of the measuring device and the
way the sample sets $U_i$ are partitioned into subsets of size
$\delta$. For this reason, different measuring devices lead to
different values of the probabilities (3.8). We may consider, for
instance two different measuring devices, one with resolution
$\delta$ and one with resolution $2 \delta$. The probabilities
corresponding to the former will be constructed from the minimal
projectors $\hat{P}^{\delta}_x$, while the latter from the
projectors $\hat{P}^{2\delta}_x$. Given that $\hat{P}^{2\delta}_x
= \hat{P}^{\delta}_{x - \frac{\delta}{2}} + \hat{P}^{\delta}_{x +
\frac{\delta}{2}}$ the difference between the probabilities
$p_{\delta}(U_i,t_1|U_j, t_2)$ and $p_{2\delta}(U_i,t_1|U_j, t_2)$
 equals
\begin{eqnarray}
\epsilon_{\delta}(U_i, t_1; U_j, t_2) = Re \, \int_{U_i} dx_1
\int_{U_j} dx_2 d_{\delta}(x_1 + \delta/2, x_1 - \delta/2, t_1:
x_2, t_2),
\end{eqnarray}
in terms of the interference term
\begin{eqnarray}
 d_{\delta}(x_1 +\delta/2, x_1 - \delta/2,t_1: x_2, t_2) =
 \nonumber \\ Tr \left(
e^{i\hat{H}(t_2-t_1)} \hat{P}^{\delta}_{x_2}
e^{-i\hat{H}(t_2-t_1)} \hat{P}^{\delta}_{x_1+\delta/2}
\hat{\rho}(t_1) \hat{P}^{\delta}_{x_1-\delta/2} \right).
\end{eqnarray}
 The probabilities for the same events depend on the resolution,
 unless  the interference term  vanishes for
all $\delta$ or if it is number of the order of
$e^{-L^2/\delta^2}$. Only then would the probabilities have a
functional dependence on the projectors $P_U$. This is the case,
for instance, when the final projector is equal to the unity, in
which case we recover the single-time results.

\subsection{An explicit example}
It is instructive to compute the probabilities and the
interference term in a concrete physical system. We assume a
non-relativistic free particle in one dimension, with Hamiltonian
$H = \frac{\hat{p}^2}{2m}$, where $m$ is the particle's mass and
$\hat{p}$ its momentum. It is convenient to employ a Gaussian
initial state
\begin{eqnarray}
\psi(x) = \frac{1}{(\pi \sigma^2)^{1/4}} e^{- \frac{x^2}{2
\sigma^2} + i p x}.
\end{eqnarray}
The parameter $\sigma$ is the position uncertainty, $p$ its mean
momentum, and we assumed without loss of generality that it is
cenered around $x=0$.

Our device measures position. We shall employ smeared Gaussians
instead of sharp projection operators \footnote{The smeared
Gaussians project, in effect, into a fuzzy set, which is quite
appropriate for realistic position measurements}.
\begin{eqnarray}
\langle x|\hat{P}^{\delta}_{x_0} |y \rangle = e^{-\frac{1}{2
\delta^2}(x - x_0)^2} \delta (x,y)
\end{eqnarray}
We are interested in the case that $ \delta << |x_0|<< \sigma$,
namely that the device may distinguish between different readings
that  lie  close to the center of the initial state. In that case
\begin{eqnarray}
\hat{P}^{\delta}_{x_0} \psi(x) = \frac{1}{(\pi \sigma^2)^{1/4}}
e^{- \frac{(x-x_0)^2}{2 \delta^2} + i p x}.
\end{eqnarray}
The above expression is, in fact, a reasonable approximation for
any wave function with spread $\sigma$, whose structure does not
vary much in the scale of $\delta$.

 Assuming that the second measurement takes place at time $t$,
 we find
 \begin{eqnarray}
p_{\delta}(x,0;x', t) &=& \pi \frac{\delta}{\sigma} \frac{r}{1+r}
e^{- \frac{a}{4 \delta^2} (x' - x - \frac{p}{m}t))^2} \\
d_{\delta}(x +\delta/2,x - \delta/2,0: x' t) &=&
p_{\delta}(x,0;x', t) e^{-c} e^{i p \delta + \frac{ib}{\delta} (x'
- x - \frac{p}{m}t)},
 \end{eqnarray}
where
\begin{eqnarray}
a &=& 1 + \frac{2 r}{(1+r)^2}, \\
b &=& 2 r \frac{(1-r)^2}{1 +r^2}, \\
c &=& \frac{r (1 +2r -3r^2 +2r^3)}{2(1+r^2)^2}, \\
r &=& \frac{m \delta^2}{t}.
\end{eqnarray}
The parameter $r$ is the time-of-flight phase space uncertainty,
namely the uncertainty $m \delta /t$ in a time-of-flight
determination of the momentum times the resolution $\delta$. From
the equations above, we see that the interference term is of the
same order as the probability irrespective of the value of $r$.

We will now estimate the difference $\epsilon$ between the
probabilities for two sets $U_1$ at time $t=0$ and $U_2$ at time
$t$. Both sets are assumed to be of size $L >> \delta$ so that the
sampling of the data can be accurate. The set $U_1$ is centered
around the point $x_1$ and the set $U_2$ around $x_2$. We will
denote by $\Delta = x_2 - x_1 - \frac{p}{m}t$ the distance between
$X_2$ and the evolution of $x_1$ according to the classical
equations of motion. Using  Gaussian smeared characteristic
functions $e^{-\frac{(x - x_i)^2}{2 L^2}}$, to perform the
integrations over the sets $U_i$, we obtain
\begin{eqnarray}
p_{\delta}(U_1,0;U_2, t) &=& k \frac{L}{\sigma} e^{-
\frac{\Delta^2}{4L^2}}, \\
\epsilon_{\delta}(U_1,0;U_2,t) &=& k' \frac{L}{\sigma} e^{-
\frac{\Delta^2}{4 L^2}} \cos \left( p \delta + \frac{b
\Delta}{\delta} \right),
\end{eqnarray}
where $k$ and $k'$ denote terms of the order of unity and
according to our original assumptions $L <<
\sigma$\footnote{Otherwise, the origin of the set $U_1$ is outside
the support of the initial wave-function, and the corresponding
probabilities will be close to zero.} . Clearly
$\epsilon_{\delta}$ is of the same order of magnitude with
$p_{\delta}$, a fact that emphasises the strong dependence of the
probabilities on the measuring device's resolution.

We conclude, therefore, that the difference between the two
probability assignments is of the same order of magnitude as the
probabilities themselves and, consequently, the probabilities for
the two-time measurements depend on the way the sample sets $U_i$
are partitioned with respect to the resolution of the measuring
device and is, therefore, not a function the sets $U_j$.

\subsection{Consequences}
Our results demonstrate that the probabilities for two-time
measurements are generically not functions of the projection
operators that correspond to the sample sets. This is very much
unlike the single time, where the probabilities for particular
events are linear functionals of the projection operators that
correspond to the sample sets.

The property above of the single-time quantum probabilities has as
consequence that all physical measurements may be described in
terms of YES-NO experiments. A typical such experiment involves a
filter --represented by a projection operator -- which is set on
the path of a particle beam. We may monitor, whether a particle
passed through the filter or not, and it is possible after $N$
trials to determine the probability corresponding to a particle
passing through that filter. If we repeat this experiment with
different filters that correspond to the same physical property
(the associated projectors commute), we may eventually reconstruct
the probability distribution for this quantity, because the single
time probabilities are additive. Hence, the probabilities
$Tr(\hat{\rho} \hat{P}_U)$ determined by the filter measurement
of, say, position in the set $U$ coincide with the probabilities
determined by the statistical analysis of all dots that were found
in the sample set $U$ in any device that records the particle
positions. For this reason, the projection operator $\hat{P}_U$
represents the proposition that the particle's position has been
measured to lie in the set $U$, irrespective of the experimental
procedure or the details of the measuring device. From the results
of the YES-NO experiments  we can unambiguously reconstruct all
probabilistic information about a physical system.
 One is led, therefore, to
the suggestion that the projection operators refer to the
properties of quantum systems --as manifested in measurements--
and that the structure of the lattice of projection operators
represents the structure of potential quantum mechanical events.
One speaks, therefore, for the {\em quantum logic} of quantum
mechanical measurements -- or of quantum mechanical properties, if
one wishes to move beyond concrete measurement situations.

In two-time measurements the situation is  different. The measured
probabilities are not functions of the single-time projectors.
They depend instead on the properties of the measuring device and
the way each sample set is resolved into minimum resolution sets.
One may still perform two-time YES-NO experiments, by directing
the particles through two successive filters. The expression (2.7)
may be employed to this experiment. But the filter measurements do
not suffice to reconstruct the two-times probability assignment to
the physical quantity they represent; the physical probabilities
for the two-time experiments are given by equation (3.8) and not
by equation (2.7). A filter measurement for position may be, for
example, realised in terms of a wall with a slit of width $L$ in
it, which represents a subset $U$ of ${\bf R}$ . This is a very
different physical system from the one we described earlier, which
records any possible position with an accuracy of $\delta$. There
is no {\em a priori} physical reason that the probability that the
particle will cross the slit will be the same with the probability
obtained from the relative frequencies of the events within $U$ in
the latter measurement. In single-time measurements they happen to
be the same, but in two-time ones they are not.

 In other
words, the YES-NO experiments do not capture all physical
information about two time measurements -- the physical
predictions depend strongly on the properties of the measuring
devices. Hence the proposition that the particle is measured at
time $t_1$ within the set $U_1$ and at time $t_2$ within the set
$U_2$ is not universally represented by the ordered pair of
projectors $(\hat{P}_{U_1}, \hat{P}_{U_2})$. It is only
represented by these projectors when the measuring device consists
of two filters -- the first with a slit corresponding to $U_1$ and
the second with a slit corresponding to $U_2$. There is,
therefore, no universality in the representation of measurement
outcomes by pairs of projection operators, with the consequence
that the interpretation of two-time measurements in terms of
quantum logic is not possible.

We may make, in fact, a stronger statement: even for single-time
measurements the interpretation in terms of quantum logic is not
possible. The proof involves {\em reductio ad absurdum}. We
represent the single-time lattice of projection operators on a
Hilbert space $H$ as $L(H)$ and assume that each measurement
outcome may be uniquely represented by an element of $L(H)$ --the
converse need not be true. Two successive measurement outcomes
should, therefore, be represented by a pair of elements of $L(H)$,
hence an element of $L(H) \times L(H)$\footnote{The lattice of
two-time measurement outcomes should contain $L(H) \times L(H)$ as
a subset --not necessarily a sublattice. Such is the case, for
instance, in Isham's scheme \cite{I94}, where the lattice of
two-time measurements is represented by the lattice of projectors
on $L(H \otimes H)$.}. We have showed that this is not the case.
Hence there exists an error in our assumptions. The statement that
two measurement outcomes are represented by a pair of arguments of
$L(H)$ is a consequence of basic principles of logical reasoning
(and a basic axiom of set theory). Unless we assume that a
two-time measurement does not correspond to two single-time
measurements\footnote{Elaborating on Bohr's interpretation of
measurement, it is possible to assume that a two-time measurement
should not be analysed conceptually into two-single time ones. Any
measurement refers to an irreducible physical set-up and a
two-time measurement should be considered as irreducible as a
single-time one. This is, however, an extreme position, which goes
much further than Bohr's explanation of the
Einstein-Podolski-Rosen (EPR) argument. The present argument is
not, whether there exist correlations between the results of two
successive measurements, but whether we may represent
--conceptually -- the results of a two-time measurement as a pair
of single time measurement. The corresponding thesis for the EPR
experiment would be to assume that there is no way to represent
separately the measurement outcomes on the two physical subsystems
-- to even consider physical observables that refer to each
subsystem. Such a position is logically consistent, but it is so
alien to the way we actually perform experiments as to be
physically untenable. The results of a two-time measurement of
position are always two readings of position. One simply cannot
refute this fact. But even if we accepted such a destructive
thesis, we would still face the fact that the outcomes of
irreducible measurements (now meaning the two-time ones) cannot be
universally represented by projection operators.}, we are forced
to conclude that the universal representation of measurement
outcomes by projection operators is not valid even in single-time
measurements. The reason it seems possible to do so, is because in
single-time measurements the interference term $d$ of equation
(2.9) always vanishes.

The conclusion above is, in a sense, complementary to many
theorems about contextuality in quantum theory  \cite{KoSp67,
Bell}. Since a given projector may be a spectral projector of two
non-commuting self-adjoint operators, one should always make a
choice of the corresponding self-adjoint operator (the context)
before employing that projector to represent a property of a
physical system \cite{Ish97}. Our results refer to concrete
measurement situations, and for this reason their conclusions are
rather stronger. Not only is it impossible to represent a
measurement outcome by a projection operator, but the observed
probabilities depend explicitly on specific properties of the
measurement device\footnote{The same conclusion holds for
approximate projectors as can be seen from the results in out
model system. Fuzzy measurements cannot save us from the generic
dependence of quantum mechanical probabilities on the measuring
apparatus.}.

There exist many interpretation schemes that employ the lattice of
projection operators to universally represent properties of
physical systems, even outside the measurement context. What we
have demonstrated here is that the universal representation of
{\em measurement outcomes} by projection operators is untenable.
It is, therefore, {\em a priori} not contradictory to assume that
projection operators refer to properties of a physical system
outside the context of measurements. We will not comment in this
paper, whether this thesis is tenable or not.  We only remark that
it is an entirely {\em ad hoc} hypothesis in the light of our
results. The quantum logic of properties of a physical systems
cannot be considered as a generalisation of a quantum logic for
measurements.

Our conclusions are independent of the interpretation for the
quantum state (whether it is objective or subjective, whether it
refers to individual systems or to ensembles). They are also
independent of the interpretation of measurement theory (whether
we employ subjective conditional probabilities, or we think that
the reduction of the wave packet is a physical process, or there
is a duality of a classical measuring device and a quantum
system).

We must distinguish the two different roles of the sample sets $U$
and the corresponding projectors -- a distinction that is not
usually made in probability theory. A sample set $U$ may represent
a physical event, if the device can not distinguish between the
elements of $U$. In that case $U$ refers to a concrete empirical
fact. It may also represent a statement about the physical system,
namely that an event has been found within the set $U$. The latter
case, however, is not a representation of a physical fact. It is
at the discretion of the experimentalist to choose the set $U$
that he will use for the sampling of its results. The physical
probabilities should, therefore, be constructed with the first
interpretation of the sample sets in mind. These probabilities
then depend then on the construction of the physical apparatus and
interaction with the measured system. They may also depend on the
initial state of the measured system: ultra-fast neutrons, for
instance, will leave a different trace on a recording material
than slow ones. The out-coming density matrix (3.2) should, in
principle, be determined by the common reasoning that is employed
in the design of experiments: treating the apparatus as classical
and the measured system as quantum. The results of such an
analysis would not yield an expression in terms of projection
operators for the minimum resolution. There exists a degree of
fuzziness in any measurement, and for this reason the natural
mathematical objects that encode the effect of minimal resolution
should be smeared projectors: positive operators with supremum
norm less than one \cite{Dav}.

The representation  of quantum mechanical measurement outcomes by
projection operators is a consequence of a principle that is often
considered as a basic axiom of quantum mechanics: The possible
outcomes in the measurement of a physical quantity represented by
a self-adjoint operator $\hat{A}$ lie  in the spectrum of
$\hat{A}$. This postulate implies that sample sets of the
measurement of $\hat{A}$ correspond to the measurable subsets of
the operator's spectrum and, due to the spectral theorem, to the
spectral projectors of $\hat{A}$. Our results suggest strongly
that this postulate may not be appropriate in quantum theory. Its
abandonment would not change any physical predictions -- an
axiomatic framework for standard quantum mechanics is still
possible in its absence. The mean values and all higher moments
for measurements of any observable may still be obtained through
the usual rules of quantum theory. The only difference is that the
spectral resolution of an operator does not necessarily correspond
to the physical resolution of the measured values of the physical
quantity. We have discussed this issue  in references \cite{An01,
An03}, to which we refer the reader for more details.

To summarise our results, if we accept that the relative
frequencies for two-time measurements converge, we inevitably
conclude that \\ \\
i. The probabilities for two specific sample sets in a two-time
measurement is not a function of the projection operators that,
supposedly, correspond to each sample set. \\ \\
ii. The YES-NO experiments do not suffice to reconstruct all
physical predictions of quantum theory for two-time
measurements.\\ \\
iii. Projection operators cannot, in general, represent properties
of a physical system. They are only relevant to the probabilities
of specific YES-NO experiments. \\ \\
iv. Unlike classical probability theory, quantum theory
distinguishes sharply between physical events and propositions
about physical events.

\section{An alternative explanation}

Our resolution of the `paradox' of two-time measurements and the
subsequent analysis was based on the assumption that the measured
frequencies of events define probabilities, i.e. that the
sequences (2.5) converge. We are then led to a reconsideration of
the use of the conditional probability for the derivation of
(2.7). The conclusion that probabilities depend rather strongly on
the properties of the physical device, is rather disturbing. It
implies that the results of two sets of measurements that involve
an identical preparation of the physical system and very similar
measuring apparatus would differ according to rather trivial
details of the apparatus's manufacture.

There exists an alternative solution to the problem, but its
implication are more disturbing rather than less. It is
conceivable that the sequences do not converge to probabilities.
They could perhaps exhibit an oscillating behavior as $N
\rightarrow \infty$. This possibility has not, to the best of our
knowledge, been refuted by any experiment that has been performed
so far. In that case probabilities cannot be defined for two-time
measurement. The expressions (2.7) cannot, therefore, be
considered as referring to probabilities. In fact, we have no idea
how to interpret them.

 The hypothesis that two-times probabilities are not defined
 is not incompatible with the successful use of
probability theory for single-time measurements. Probabilities are
additive for single-time measurement -- just as relative
frequencies are-- and there is no problem in that case to assume
that the sequences (2.5) converge. The same would be true for any
sufficiently coarse-grained measurements, for which the
interference term vanishes. One would be, therefore, led to the
interpretation of the object $| Re \,d(U_1,U_j;U_2,U_j)|$ as a
measure of the non-convergence of the sequence of relative
frequencies. Probabilities would, therefore, be definable only for
specific samplings of the measurement outcomes, such that the
consistency condition $Re \, d(U_1,U_j;U_2,U_j) = 0$ holds. This
is, in fact, similar to the use of probabilities by the consistent
histories approach -- probabilities are defined only for
sufficiently coarse-grained partitions of the two-time sample
space, such that the consistency condition is satisfied.

The second alternative is perhaps too radical. It would involve a
reappraisal of the use of probabilities in physical theories. We
would have to extend both the theory of probability and quantum
theory in a way that will deal with non-convergent sequences of
relative frequencies. It is not clear, how this may be achieved,
what generalisations are physically relevant, and which parts of
the quantum mechanical formalism, if any,  would have to be
abandoned. As far as the explanation of the non-convergence of
frequencies is concerned, we may only speculate. Perhaps, the
set-up of two time measurements does not lead to the probabilities
of  quantum equilibrium -- in the sense of Bohmian mechanics
\cite{BH}. Or, perhaps, it is due to a physical reason unsuspected
by any current interpretation or reformulation of quantum theory.
In absence of conclusive empirical evidence, we find more prudent
to refrain from  any detailed speculation on this issue.

The hypothesis of non-converging frequencies  seems much less
plausible than the alternative we considered in the previous
sections. However, it is {\em a priori} possible that physical
phenomena cannot be entirely described in terms of probabilities.
In any case, this issue can be resolved by recourse to experiment.
It should not be very difficult to design and execute experiments
that will measure
 particle positions at two moments of time. A careful
 statistical analysis of the measurement outcomes will then allow
 us to clearly distinguish whether the relative frequencies
 converge or not. If they do not, then it would be a strong
 argument in support of the incompleteness of the current
 formulation of quantum theory.

\section*{Acknowledgements}
I would like to thank N. Savvidou for many discussions and
comments.

\end{document}